\documentclass[prd,aps,twocolumn]{revtex4}
\usepackage{latexsym}
\usepackage{amssymb}
\usepackage[dvips]{graphicx}
\begin{document}
\newcommand{\beq}{\begin{equation}}
\newcommand{\eeq}{\end{equation}}
\newcommand{\ove}{\overline}
\newcommand{\half}{\frac 1 2 }
\newcommand{\fourth}{\frac 1 4}
\newcommand{\Fstar}{\raisebox{.2ex}{$\stackrel{*}{F}$}{}}
\newcommand{\Pstar}{\raisebox{.2ex}{$\stackrel{*}{P}$}{}}
%
\newcommand{\et}{{\em et al}}
\newcommand{\ie}{{\em i.e.$\;$}}
%
%
\newcommand{\Prd}{Phys.  Rev. D$\;$}
\newcommand{\Prl}{Phys.  Rev.  Lett.}
\newcommand{\Plb}{Phys.  Lett.  B}
\newcommand{\Cqg}{Class.  Quantum Grav.}
\newcommand{\Np}{Nuc.  Phys.}
\newcommand{\Grg}{Gen.  Rel.  \& Grav.$\;$}
\newcommand{\Fp}{Fortschr.  Phys.}
\newcommand{\Sch}{Schwarszchild$\:$}
\renewcommand{\baselinestretch}{1.2}

\title{Constraining $f(R)$ theories with the energy conditions}

\author{S. E. Perez Bergliaffa}
\altaffiliation[On leave from ]{\emph{Departamento de F\'{\i}sica Te\'{o}rica,
Instituto de F\'{\i}sica, Universidade do Estado de Rio de Janeiro,
CEP 20550-013, Rio de Janeiro, Brazil}.}
\affiliation{ICRANET, Piazzale della Repubblica, 10 - 65100 Pescara, Italia}

\vspace{.5cm}

\begin{abstract}
A new method to constrain gravitational theories depending on the Ricci scalar
is presented. It is based on the weak energy condition and yields limits on the parameters of a given theory
through the current values of the derivatives of the scale factor of the Friedmann-Robertson-Walker geometry.
A further constraint depending on
the current value of the snap is also given. Actual constraints (and the corresponding error
propagation analysis) are
calculated for two examples, which show that the method is useful in limiting the possible $f(R)$
theories.
\end{abstract}
\date{\today}

\vskip2pc
\maketitle

\section*{Introduction}

It follows from several observations \cite{obsaccun} that the universe is
currently expanding with positive acceleration. The many models
that have been advanced to explain this situation can be classified in
two classes. The first class contains those models that incorporate
modifications to the matter side of Einstein's equations. This
matter (known as ``dark energy"), can be described by an ideal fluid or by a scalar
field with a convenient potential \cite{ratra}, and it must violate the
strong energy condition in order to accelerate the universe in
General Relativity (GR). The models in the second class
have normal matter as a source but assume that gravitation is not described by GR at low
curvatures. As examples of this latter class we can mention theories depending
on the Ricci scalar
\cite{carroll, gra} (the so-called $f(R)$ theories), and gravity modified from contributions of
extra dimensions \cite{ced}.
Although $f(R)$
theories
offer a chance to explain the acceleration of the universe, they are not free
of
problems. For instance, the application of their metric formulation
to the
Friedmann-Robertson-Walker (FRW) geometry yields a
fourth order nonlinear differential equation for the scale factor
$a(t)$ which in general cannot be analytically solved even for
simple $f(R)$.

Since many $f(R)$ give rise to accelerated expansion, another
issue is how to reduce the theory space
using observations.
Constrains have been obtained from cosmological and astrophysical
data, solar system tests, fifth force/BBN data, and by requiring that a given theory
describes the correct sequence of decelerated-accelerated phases in
the evolution of the universe \cite{odi}. Most of the cosmological tests
involve either some transformation of the theory under scrutiny to an equivalent form with one auxiliary scalar field
and/or considerations in different frames
(see for instance \cite{broo}),
or some assumptions regarding the dependence of the Hubble
``constant" $H$ with the redshift
\cite{capon}. Here instead a new criterion based on model-independent data
shall be given,
that helps in deciding, without solving the EOM or making frame transformations or assumptions about $H$,
whether a given $f(R)$ theory is appropriate to describe the universe. The basic premise will be that
the acceleration
is due solely to a modified gravitational theory with normal matter as a source.
The criterion
will then be obtained by  imposing the energy conditions on matter, yielding
conditions on $f(R)$ and its derivatives w.r.t $R$ in terms of the current value of
the derivatives of the scale factor. These conditions are to be satisfied if the theory given by
$f(R)$ is to describe the current state of the universe, and they bring forth
limits on the parameters that enter the theory
under consideration.

\section*{Energy Conditions}

The energy conditions (EC) are inequalities satisfied by ``normal" matter
(see for instance \cite{mattbook}).
When specialized to a FRW universe, the (local) null, weak, strong and dominant EC are given by
$$
{\rm NEC}\Longleftrightarrow \rho + p \geq 0,
$$
$$
{\rm WEC} \Longleftrightarrow \rho\geq 0\,\, {\rm and}\,\, (\rho + p \geq 0),
$$
$$
{\rm SEC} \Longleftrightarrow (\rho + 3p \geq 0)\,\, {\rm and}\,\,(\rho + p \geq 0),
$$
$$
{\rm DEC} \Longleftrightarrow \rho\geq 0\,\, {\rm and}\,\, (\rho\pm p\geq 0).
$$
The EC have proved to be useful in the context of
cosmological singularities \cite{haw} and bounces \cite{matt}. Other
applications of the energy conditions to cosmology can be found in
\cite{mattscience}.

Let us remind the reader that for a theory given by $f(R)$, the EOM
are \cite{kerner}
\beq f'R-2f+3f''\left(\ddot R + \frac{3\dot a \dot
R}{a}\right)+3f'''\dot R^2+ T=0,
\label{k1}
\eeq
\beq f'R_{tt}+\half
f-3f''\frac{\dot a\dot R}{a}+ T_{tt}=0,
\label{k2}
\eeq
where $f'\equiv\frac{df}{dR}$, etc,
$$
R­_{tt}=\frac{3\dot a}{a},\,\,\,\,\,\,\,\,\,\,\,\,R=-6\left(\frac{\ddot a}{a} + \frac{\dot a^2}{a^2}\right),
$$
and we have assumed a flat universe.
From Eqns.(\ref{k1}) and (\ref{k2}) the energy density and the pressure of the
fluid can be expressed in terms of the scale factor and its derivatives:
\beq
\rho=-f' R_{tt}-\frac{f}{2}+3f''\frac{\dot a \dot R}{a},
\label{rho}
\eeq
\beq
p=-\frac{f'}{3}\left(R_{tt}+R\right)+\frac f 2 -f'' \left( \ddot
R-\frac{2\dot a \dot R}{a}\right)-f'''\dot R^2.
\label{p}
\eeq
Before proceeding to build with these equations the inequalities that define the energy conditions,
let us remark that observations show that the current matter content of
the universe
(assumed here to be normal matter, as opposed to dark energy) is pressureless. In this case the EC reduce to
the inequality
$\rho_0\geq 0$ plus the equation $p_0=0$, where the subindex 0 means that the quantity is evaluated today.
We shall express these conditions in terms of following kinematical parameters:
the Hubble and deceleration parameters, the jerk, and the snap,
respectively given by \cite{mattcqg}
$$
H=\frac{\dot a}{a},\,\,\,\,\,\,\,\,\,\,\,\,\,\,\,q=-\frac{1}{H^2}\frac{\ddot a}{a},
$$
$$
j=\frac{1}{H^3}\frac{\stackrel{...}{a}}{a},\,\,\,\,\,\,\,\,\,\,\,\,\,\,\,s=\frac{1}{H^4}\frac{\stackrel{....}{a}}{a}.
$$
While the current value of the first three parameters is
$
H_0=72\pm 8\,{\rm km/(secMpc)} $ \cite{free},
$q_0=-0.81\pm 0.14$, $j_0=2.16^{+0.81}_{-0.75}$
\cite{blan},
no measurements of the snap have been reported yet. By writing
$\rho_0\geq 0$ in terms of the parameters we get
\beq
3q_0H_0^2f_0'-\frac{f_0}{2} -18 H_0^4f_0''(j_0-q_0-2)\geq 0.
\label{r1}
\eeq
This inequality gives a relation between the derivatives of a given $f(R)$ and, as will be seen in the examples of the
next section, it limits the possible values of the parameters of the theory.

Notice that Eqn.(\ref{p}) involves the snap (through $\ddot R$). If we had a measurement
of $s_0$, we could use the equation
$p_0=0$ to obtain another constraint on $f(R)$. Since this is not the case, we shall express $p_0=0$ in such a way
that it gives
the possible current values of the snap
for a given $f(R)$:
\begin{eqnarray}
s_0 & = & \frac{f_0'}{6H_0^2f_0''}(q_0-2)+
 6H_0^2\frac{f_0'''}{f_0''}(-q_0+j_0-2)^ 2-  \nonumber \\
 & &
[q_0(q_0+6)+2(1+j_0)] -\frac{f_0}{12H^4f_0''}.
\label{r2}
\end{eqnarray}

\section*{Examples}

\begin{figure}[h]
\begin{center}
\includegraphics[width=0.45\textwidth]{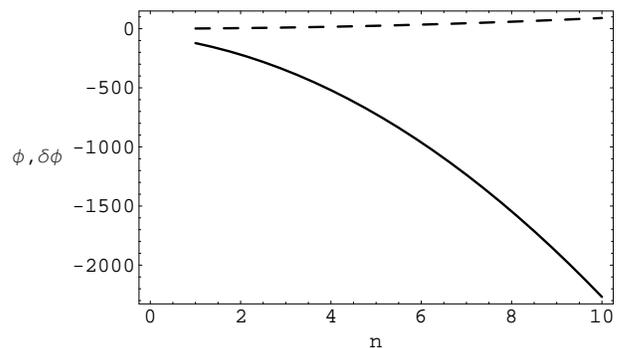}
\caption{Plot of $\phi$ (solid curve, see Eqn.(\ref{neqn})) and the associated error $\delta\phi$
(dashed curve)
in terms of $n$.}
\label{ex1}
\end{center}
\end{figure}

\begin{figure}[h]
\begin{center}
\includegraphics[width=0.45\textwidth]{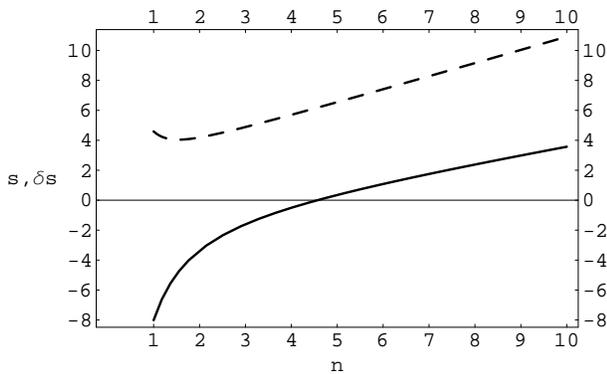}
\caption{Plot of the snap for $f(R)$ given in Eqn.(\ref{f1})
(solid curve) and the associated error (dashed curve)
in terms of $n$.}
\label{snap}
\end{center}
\end{figure}

To see how Eqns.(\ref{r1}) and (\ref{r2}) can be used to put constraints on a given $f(R)$,
let us examine two examples. The first one is given by
\cite{capo}
\beq
f(R)=\alpha R^{-n},
\label{f1}
\eeq
with $n\in \mathcal{N}$,
which can be taken as the $n=1$ low-curvature limit of
$$
f(R)=R+\frac\alpha R,
$$
a model studied in \cite{carroll}. Substituting Eqn.(\ref{f1}) in
Eqn.(\ref{r1}) for $\alpha >0$ and $n$ even we get
\beq
-3q_0H_0^2nR_0-\half R_0^2 -18H_0^4n(n+1)(j_0-q_0-2)\geq 0.
\eeq
Replacing in this equation the numerical values of the parameters
and using $R_0=6H_0^2(q_0-1)$ we get an inequality that must be
satisfied by $n$:
\beq \phi \equiv -17.64n^2-44.50n-59.62\geq 0.
\label{neqn}
\eeq
Since this equation cannot satisfied by any real $n$, we
conclude that $n$ cannot be even for $\alpha >0$. The same analysis with odd $n$ reverses the sign of the inequality
(\ref{neqn}), so only odd values of $n$ are allowed for $\alpha >0$. This result
generalizes that obtained in \cite{carroll} for $n=1$ \footnote{Note
the difference in the sign of $R$ between our convention and the one
used in \cite{carroll}.}. In the same way, we obtain that only even
$n$ are allowed for $\alpha <0$. These conclusions are valid even
when the error coming from the kinematical parameters is taken into
account (see Fig.\ref{ex1}).
If we knew the value of $s_0$, we could get a further constraint for the possible values of $n$
using Eqn.(\ref{r2}).
This equation will be taken instead as giving the current value of the
snap as a function of $n$ (see fig.(\ref{snap}))\footnote{Note that Eqn.(\ref{r2})
is independent of the sign of
$\alpha$.}.
We have also plotted in the figure the error associated with $s_0$ (dashed line), which grows as $n$ for large $n$.
The plot shows that with the current error of the kinematical parameters the method outlined here is helpful in
determining $s_0$ for the theory given by Eqn.(\ref{f1})
only for $n=1$.

\begin{figure}[h]
\begin{center}
\includegraphics[width=0.45\textwidth]{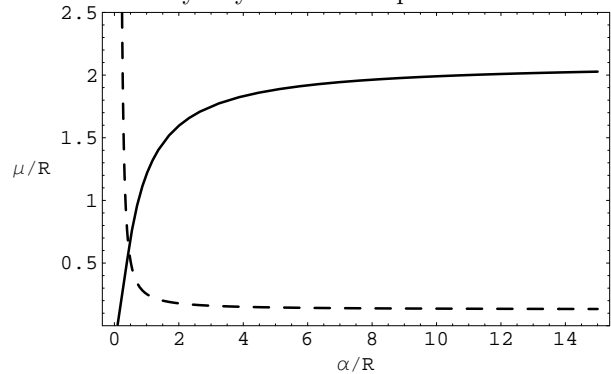}
\caption{Plot of $\mu/R_0$ (solid curve) and the associated
error (dashed curve) in terms of $\alpha/R_0$. The allowed values for $\mu$ are those below the curve.}
\label{mu}
\end{center}
\end{figure}

Let us next analyze an example of a theory involving two dimension-full parameters, given by \cite{noji}
\beq
f(R)=R+\alpha \ln \frac R \mu,
\label{ji}
\eeq
where $\mu<0$. In this case it follows from Eqn.(\ref{r1}) and $\alpha<0$
that
\beq
0<\frac{\mu}{R_0} <e^{-g(\beta)}
\eeq
where $\beta=\alpha/R_0$ and
\beq
g(\beta) =\frac 1 \beta \left[-6q_0(1+\beta)\frac{H_0^2}{R_0}+1-36\frac{H_0^4}{R_0^2}\beta (j_0-q_0-2)\right].
\eeq
Figure (\ref{mu}) shows the permitted values for $\mu/R_0$ in the case $\alpha<0$,
which are between the horizontal axis and the solid curve,
as well as the associated error
\footnote{For $\alpha>0$, the values below the solid curve are the excluded ones.}.
The plot shows that $\mu/R_0$ tends to a constant value ($\approx 2.1$) for large $\beta$. Hence, the
possible values of $\mu$ are restricted to $|\mu| \lesssim 1.2\times 10^{-41}$ m$^{-2}$.

In the case of the theory given in Eqn.(\ref{ji}), the
snap would be a function of $\alpha$, $\mu$, and of the
remaining kinematical parameters.

\section*{Discussion}

A new method to restrict gravitational
theories described by
functions of the Ricci scalar has been introduced.
It is based essentially in the assumption that normal matter composes
the universe, the acceleration being caused by new gravitational dynamics in the low curvature regime,
described by $f(R)$.
By imposing that the matter satisfy the weak energy condition, we obtain an inequality that
constrain the parameters in the theory. We have shown by way of two examples how the method can be used,
and how it conduces to restrictive limits on the parameters, having taken the error into account.
We also obtained an equation that depends on the snap, the fourth derivative of the scale factor.
Had we any measurements of $s_0$, this equation would furnish yet another condition on the parameters of the
theory. Since the current value of the snap has not been determined yet, we take this equation as
forecasting, for a given $f(R)$, the current value of the snap.
The method presented here could be combined with other approaches (such as avoidance of super-luminal propagation
speed \cite{defe}, compatibility with the PPN limit \cite{sta}, or those mentioned in the introduction)
to restrict the $f(R)$ theories that are being presented as candidates to model the
acceleration of the universe.

\section*{Acknowledgements}

The author would like to thank ICRANET for financial support and hospitality
during the preparation of this manuscript.

\end{document}